\documentclass[aps,prd,superscriptaddress,nofootinbib,floats,preprint,floats,floatfix]{revtex4-1}
\usepackage{bm}
\usepackage{indentfirst}
\usepackage{amsmath}
\usepackage{graphicx}
\usepackage{float}
\usepackage{amssymb}
\usepackage{subfigure}
\usepackage{psfrag}
\usepackage{hyperref}
\usepackage{tensor}
\usepackage{braket}
\usepackage{multirow}
\usepackage{verbatim}
\hypersetup{
	colorlinks=true,
	linkcolor=red,
	citecolor=blue,
} 
\allowdisplaybreaks[1]

\begin{document}
\title{Circularly polarized scalar induced gravitational waves from the Chern-Simons modified gravity}
\author{Fengge Zhang}
\email{zhangfg5@mail.sysu.edu.cn}
\affiliation{School of Physics and Astronomy, Sun Yat-sen University, Zhuhai 519088, China}
\author{Jia-Xi Feng}
\email{fengjx57@mail2.sysu.edu.cn}
\affiliation{School of Physics and Astronomy, Sun Yat-sen University, Zhuhai 519088, China}
\author{Xian Gao}
\email{gaoxian@mail.sysu.edu.cn (corresponding author)}
\affiliation{School of Physics and Astronomy, Sun Yat-sen University, Zhuhai 519088, China}

\begin{abstract}
We investigate the scalar induced gravitational waves (SIGWs) in the Chern-Simons (CS) modified gravity during the radiation dominated era. The SIGWs are circularly polarized, which provide us a tool to test the possible parity violation in the early universe. We derive the semianalytic expressions to  evaluate the fractional energy density of the SIGWs, $\Omega_{\mathrm{GW}}$, which  receives contributions from the general relativity (GR)  and the correction due to the parity-violating term, respectively. 
We find that the degree of the circular polarization of the SIGWs can be as large as of order unity, although the contribution to $\Omega_{\mathrm{GW}}$ from the CS term is at most of the same order as that from the GR.
\end{abstract}

\maketitle

\section{Introduction}

The detection of the gravitational waves (GWs) by the Laser Interferometer Gravitational-Wave Observatory (LIGO) scientific collaboration and Virgo collaboration \cite{Abbott:2016nmj,Abbott:2016blz,Abbott:2017gyy,TheLIGOScientific:2017qsa,Abbott:2017oio,Abbott:2017vtc,LIGOScientific:2018mvr,Abbott:2020khf,Abbott:2020uma,LIGOScientific:2020stg} opens a new window to probe the nature of gravity. Although a variety of information in understanding the early universe and fundamental physics is encoded in the primordial GWs, the primordial GWs have not been detected on the cosmic microwave background (CMB) scale  up to now \cite{Akrami:2018odb,BICEP:2021xfz}. On the other hand, the scalar induced gravitational waves (SIGWs) \cite{Ananda:2006af,Saito:2008jc,Orlofsky:2016vbd,Nakama:2016gzw,Wang:2016ana,Cai:2018dig,Kohri:2018awv,Espinosa:2018eve,Kuroyanagi:2018csn,Domenech:2019quo,Fumagalli:2020nvq,Lin:2020goi,Domenech:2020kqm,Domenech:2021ztg,Zhang:2021vak,Wang:2021djr,Adshead:2021hnm,Ahmed:2021ucx,Zhang:2021rqs}, which are sourced by the first order scalar perturbations due to the non-linearity of gravity,  have attracted much attention recently. The SIGWs can be large enough to be detected by the space-based GW observatories such as  Laser Interferometer Space Antenna (LISA) \cite{Danzmann:1997hm,LISA:2017pwj}, TianQin \cite{Luo:2015ght} and Taiji \cite{Hu:2017mde}, as well as by the Pulsar Timing Array (PTA) \cite{Kramer:2013kea,Hobbs:2009yy,McLaughlin:2013ira,Hobbs:2013aka} and the Square Kilometer Array (SKA) \cite{Moore:2014lga} in the future. This is due to the fact that the amplitude of the power spectrum of the primordial curvature perturbation is $\mathcal A_{\zeta}\sim \mathcal{O}(10^{-2})$ on small scales \cite{Sato-Polito:2019hws,Lu:2019sti,Khlopov:2008qy}, which can be  seven orders of magnitude larger than the constraint from the CMB, that is, $\mathcal A_{\zeta}\sim\mathcal{O}(10^{-9})$ on large scales \cite{Akrami:2018odb}. Recently, North American Nanohertz Observatory for Gravitational Waves (NANOGrav) 12.5 yrs data also indicates that we may have captured the SIGW signals \cite{DeLuca:2020agl,Vaskonen:2020lbd,Kohri:2020qqd,Domenech:2020ers}.

In light of these progresses, in this work we use the SIGWs to test the possible parity violation in the early universe.
Such parity-violating (PV) terms are generally predicted in quantum gravity theories such as the superstring theory and M-theory \cite{Green:1984sg,Witten:1984dg}. 
One of the PV terms is the Chern-Simons (CS) modified gravity, which was first proposed in four-dimensional space-time in \cite{Jackiw:2003pm}, and then
has been studied extensively in cosmology and GWs \cite{Lue:1998mq,Satoh:2007gn,Saito:2007kt,Satoh:2007gn,Alexander:2009tp,Yunes:2010yf,Gluscevic:2010vv,Yunes:2010yf,Myung:2014jha,Kawai:2017kqt,Nair:2019iur,Nishizawa:2018srh,Odintsov:2019mlf,Odintsov:2022hxu,Odintsov:2022cbm}. The non-Gaussianity in CS gravity is also studied in \cite{Bartolo:2017szm,Bartolo:2018elp}. Moreover, the CS gravity propagates Ostrogradsky ghosts due to the existence of higher-derivate field equation, it can only be treated as a low-enery truncation of a fundamental theory. 
Since the cosmological background breaks the Lorentz symmetry, the PV gravity models with Lorentz breaking, such as Ho\v{r}ava gravity \cite{Horava:2009uw}, the PV higher derivative gravity \cite{Crisostomi:2017ugk} and the PV spatially covariant gravity \cite{Gao:2019liu,Hu:2021bbo,Hu:2021yaq} have also been proposed.
The chiral GWs in such Lorentz breaking PV gravity models have been studied in \cite{Takahashi:2009wc,Myung:2009ug,Wang:2012fi,Zhu:2013fja,Cannone:2015rra,Zhao:2019szi,Zhao:2019xmm,Qiao:2019hkz,Qiao:2019wsh,Qiao:2021fwi,Gong:2021jgg}.
When torsion is turned on, the simplest term that corresponds to the CS term is the so-called Nieh-Yan term \cite{Nieh:1981ww}. 
The chiral GWs have also been extensively studied with NY term and its extensions  \cite{Chatzistavrakidis:2020wum,Cai:2021uup,Wu:2021ndf,Langvik:2020nrs,Li:2020xjt,Li:2021wij,Rao:2021azn,Li:2021mdp,Li:2022mti,Li:2022vtn}, as well as in more general models with non-vanishing torsion and/or non-metricity tensors  \cite{Hohmann:2020dgy,Bombacigno:2021bpk,Iosifidis:2020dck,Hohmann:2022wrk,Conroy:2019ibo,Iosifidis:2021bad,Pagani:2015ema}.
To our knowledge, the previous studies with the PV modes dealt with only the primordial/linear GWs, which exhibit interesting features such as the velocity and amplitude birefringence phenomenons of GWs \cite{Takahashi:2009wc,Wang:2012fi,Alexander:2004wk,Yunes:2010yf}, see also \cite{Wang:2020pgu,Wang:2021gqm,Wang:2020cub,Zhao:2022pun}.
The PV effects on the SIGWs have been less investigated.

In this work we make the first step to use the SIGWs to test such PV phenomena. Usually the behaviour of matter contents of the universe on cosmological scales is mimicked by either perfect fluid or scalar condensation. 
In this paper we focus on the simplest CS gravity coupled with a dynamical scalar field, which effectively describes the matter contents of the universe.
In particular, we assume the equation of state of the scalar field to be the form of the radiation, which enables us to investigate the implications to the SIGWs during the radiation dominated era.
Moreover, we choose the coupling function to be the exponential form, so that we can obtain an analytic expression for the Green's function that is used to solve the equation of motion of the SIGWs.
We give the semianalytic expression used to calculate the power spectrum of SIGWs from the CS modified gravity, which contains the contributions from the general relativity (GR), $I_{\mathrm{GR}}$ and from the PV term, $I^A_{\mathrm{PV}}$, respectively. In order to study the features of SIGWs from CS modified gravity, we evaluate the fractional energy density $\Omega_{\mathrm{GW}}$ and the degree of the circular polarization of SIGWs with the monochromatic and lognormal power spectra of primordial curvature perturbation, respectively.

This paper is organized as follows. In section \ref{II}, we briefly introduce the  CS modified gravity model and the equations of motion.  In section  \ref{III}, we first derive the equations of motion for the SIGWs from CS gravity, and then give the semianalytic expressions used to compute the power spectrum of the SIGWs. In section \ref{IV}, we compute the fractional energy density and the degree of the circular polarization of SIGWs from CS modified gravity with some phenomenological examples to study the features of SIGWs. 
Our main conclusions are summarized in section \ref{V}. 
We give the detailed calculations and the tedious semianalytic expressions in appendix \ref{kernelI}.

\section{Chern-Simons modified gravity}\label{II}

In this work, we consider the simplest Chern-Simons (CS) gravity coupled with a dynamical scalar field, of which the action is given by
\begin{equation}\label{action}
S=\frac{1}{2\kappa^2}\int \mathrm{d}^4x \sqrt{-g}~[R+\frac{1}{4}\vartheta(\varphi) {}^{\ast}RR~]-\int \mathrm{d}^4x\sqrt{-g}\left(\frac{1}{2}g^{\mu \nu}\nabla_{\mu}\varphi\nabla_{\nu}\varphi +V(\varphi)\right),
\end{equation}
where $\kappa^2=8\pi G$, and ${}^{\ast}RR$ is the CS term
\begin{equation}
{}^{\ast}RR= {}^{\ast}R^{\mu \nu \rho \gamma}R_{\mu \nu \rho \gamma}.
\end{equation}
The dual Riemann tensor is defined by
\begin{equation}
{}^{\ast}R^{\mu \nu \rho \gamma}=\frac{1}{2}\varepsilon^{\rho \gamma \alpha \beta}R^{\mu \nu}_{\ \ \ \alpha \beta},
\end{equation}
with $\varepsilon^{\rho \gamma \alpha \beta}$ the Levi-Civita tensor.

By varying the action with respect to the metric tensor $g_{\mu \nu}$ and scalar field $\varphi$, we obtain the equations of motion (EoMs) of CS modified gravity, which are
\begin{equation}\label{eqg}
G_{\mu \nu}+C_{\mu \nu}=\kappa^2 T_{\mu \nu}, 
\end{equation}
\begin{equation}\label{eqm}
g^{\mu\nu}\nabla_{\mu}\nabla_{\nu}\varphi - V_{\varphi}(\varphi)+\frac{\vartheta_{\varphi}}{8\kappa^2} {}^{\ast}RR=0,
\end{equation}
where $\vartheta_\varphi=d\vartheta/d\varphi$, $V_{\varphi}=dV/d\varphi$, and
\begin{gather}
G_{\mu \nu}=R_{\mu \nu}-\frac{1}{2}g_{\mu \nu}R, \\ 
C_{\mu \nu}=\nabla_{\alpha}\vartheta \varepsilon^{\alpha \beta}_{\ \ \gamma(\mu}\nabla^{\gamma}R_{\nu)\beta}+\nabla_{\alpha}\nabla_{\beta}\vartheta {}^{\ast}R^{\beta\ \ \ \alpha}_{\ (\mu \nu)},\\
T_{\mu \nu}=\nabla_{\mu}\varphi \nabla_{\nu}\varphi-g_{\mu \nu}\left(\frac{1}{2}g^{\alpha \beta}\nabla_{\alpha}\varphi\nabla_{\beta}\varphi+V(\varphi)\right).
\end{gather}
In the above, $G_{\mu\nu}$ is the Einstein tensor, $C_{\mu\nu}$ encodes the contribution from the CS term.
The energy-momentum tensor $T_{\mu\nu}$ can be recast to be in the form of the perfect fluid
\begin{equation}
    T_{\mu \nu}=(\rho+P)U_\mu U_\nu +g_{\mu \nu}P,
\end{equation}
with
\begin{gather}
U_\mu=\frac{\nabla_\mu\varphi}{\sqrt{-g^{\mu\nu}\nabla_\mu \varphi \nabla_\nu \varphi}},\\
\label{rho}
\rho=-\frac{1}{2}g^{\mu \nu}\nabla_{\mu}\varphi \nabla_{\nu} \varphi+V(\varphi),\\
\label{pr}
P=-\frac{1}{2}g^{\mu \nu}\nabla_{\mu}\varphi \nabla_{\nu} \varphi-V(\varphi).
\end{gather}

\section{Scalar induced gravitational waves}\label{III}

In this section, we first derive the EoMs of the first order scalar perturbations, which are necessary to evaluate the SIGWs. The perturbed metric under the Newtonian gauge is \footnote{On the gauge dependence of SIGWs, please refer to \cite{Lu:2020diy,Ali:2020sfw,Chang:2020tji,Chang:2020iji,Chang:2020mky,Domenech:2020xin,Inomata:2019yww,Tomikawa:2019tvi,Yuan:2019fwv,DeLuca:2019ufz}.}
\begin{equation}
\mathrm{d}s^{2}=a^2\left[-(1+2 \phi) \mathrm{d} \eta^{2}+\left((1-2\psi)\delta_{i j}+\frac12 h_{ij}\right) \mathrm{d} x^{i} \mathrm{d} x^{j}\right],
\end{equation}
where $\phi, \psi$ are the first order scalar perturbations.  The tensor perturbations $h_{ij}$ are assumed to be of the second order, which are transverse and trace-free, ${\partial_{i}h_{ij}=h_{ii}=0}$. Here we do not consider the first order tensor perturbations, namely the primordial or linear GWs,  which are irrelevant to this paper. 

Expanding Eqs. \eqref{eqg} and \eqref{eqm} to the zeroth order, we obtain the background EoMs as follows,
\begin{gather}\label{BEQ}
3\mathcal{H}^2=\kappa^2 a^2 \bar{\rho},
\\ -(\mathcal{H}^2+2\mathcal{H}')=\kappa^2 a^2 \bar{P},
\\\varphi''+2\mathcal{H}\varphi'+a^2V_{\varphi}(\varphi)=0,
\end{gather}
where $\mathcal{H}=a'/a$, $\bar{\rho}$ and $\bar{P}$ represent the background energy density and pressure, respectively. The prime denotes derivative with respect to the conformal time $\eta$ defined as $\mathrm{d}\eta=\mathrm{d}t/a$. 
These background EoMs are exactly the same as those in GR, since the PV terms have no contribution at the background level.

During the radiation dominated era, the equation of state is
\begin{equation}\label{w}
w=\frac{\bar P}{\bar \rho}=\frac{1}{3}.
\end{equation}
Combining Eqs. \eqref{rho}, \eqref{pr}, \eqref{w} and the background equations, we yield the evolution of the background quantities as
\begin{equation}\label{beq}
\bar{\rho}=\rho_0 a^{-4},\ \ a=\sqrt{\frac{\kappa^2}{3}\rho_0}\eta=a_0\eta, \ \ \varphi'=\pm\frac{2}{\kappa}\eta^{-1}.
\end{equation}

At the first order, we get the EoMs of linear scalar perturbations,
\begin{gather}
 \label{1st_eq_1}
 3\mathcal{H}(\psi^\prime+\mathcal{H}\phi)-\nabla^2\psi
  =-\frac{\kappa^2}{2}a^2\delta\rho,\\
\label{1st_eq_2}
\psi^\prime+\mathcal{H}\phi=-\frac{\kappa^2}{2}a^2(\bar{\rho}+\bar{P})\delta \mathcal{U},\\
\label{1st_eq_3}
\psi-\phi=0,\\
\label{1st_eq_4}
\psi^{\prime\prime}+2\mathcal{H}
\psi^\prime+\mathcal{H}\phi^\prime+(2\mathcal{H}^\prime+\mathcal{H}^2)\phi=\frac{\kappa^2}{2}a^2\delta P,
\end{gather}
where 
\begin{gather}
    \delta \rho=-a^{-2}\varphi'(\phi \varphi'-\delta \varphi')+V_{\varphi}\delta \varphi,\\
    \delta P=-V_{\varphi}\delta \varphi +a^{-2}\varphi'(\delta \varphi' - \phi \varphi'),\\
    \label{du}
    \delta \mathcal{U}=-\frac{\delta \varphi}{\varphi'}. 
\end{gather}
Combining the background EoMs, Eqs. \eqref{1st_eq_2} and \eqref{du}, we obtain the perturbation of scalar field in terms of the scalar metric perturbations
\begin{equation}\label{pvarphi}
    \delta \varphi =\frac{\psi'+\mathcal{H}\phi}{\mathcal{H}^2-\mathcal{H}'}\varphi'.
\end{equation}
From the above equations, we can see that  the linear scalar perturbations are unaffected by PV terms either.

Expanding Eq. \eqref{eqg} to the second order and extracting its transverse and trace-free part, we obtain the EoMs for the SIGWs,
\begin{equation}\label{sgweq}
h^{''}_{ij}+2\mathcal{H}h^{'}_{ij}-\nabla^2 h_{ij}+\left[\frac{\epsilon_{ilk}}{2a^2}(\vartheta''\partial^l h_{j}^{'k}+\vartheta'\partial^{l}h_{j}^{''k}-\vartheta'\partial^{l}\nabla^2h_{j}^{\ k})+i\leftrightarrow j\right]=4\mathcal{T}^{lm}_{\phantom{lm}ij}s_{lm},
\end{equation}
where $\mathcal{T}^{lm}_{\phantom{lm}ij}$ is the projection tensor used to extract the transverse and trace-free part of a tensor, and $s_{ij}$ is
the source of SIGWs,
which consists of the scalar perturbations at the second order. As expected, $s_{ij}$ can be split into two parts,
\begin{equation}
    s_{ij}=s^{\mathrm{scalar}}_{ij}+s^{\mathrm{PV}}_{ij},
\end{equation}
where the contribution from the parity-preserving part is the same as in the GR,
\begin{equation}
\begin{split}
s^{\mathrm{scalar}}_{ij}=&-\partial_i \phi \partial_j \phi+\partial_i\psi\partial_j \phi+\partial_i \phi \partial_j \psi-3\partial_i \psi \partial_j \psi \\ &-2\phi\partial_{i}\partial_{j}\phi -2\psi \partial_i \partial_j \psi +8\pi G a^2(\bar{\rho}+\bar{P})\partial_i \delta \mathcal{U} \partial _j \delta \mathcal{U},
\end{split}
\end{equation}
with $\delta \mathcal{U}$ given in Eq. \eqref{du},
and the contribution from the PV term is
\begin{equation}
s^{\mathrm{PV}}_{ij}=\frac{1}{2a^2}\left\{\left[\epsilon_{ilk}(\vartheta_\varphi \partial^k \delta \varphi' \partial^l\partial_j\phi+\vartheta_\varphi \partial^k \delta \varphi \partial^l\partial_j\phi'+\vartheta_{\varphi\varphi}\varphi' \partial^k \delta \varphi \partial^l\partial_j\phi)+i\leftrightarrow j\right]+(\phi \rightarrow \psi)\right\}.
\end{equation}

We decompose $h_{ij}$ into circularly polarized modes as
\begin{equation}
    h_{ij}(\bm{x},\eta)=\sum\limits_{A=R,L}\int \frac{\mathrm{d}^3k}{(2\pi)^{3/2}}e^{i\bm{k}\cdot\bm{x}}p^{A}_{ij}h^A_{k}(\eta),
\end{equation}
where the circular polarization tensors are
\begin{equation}\label{cpt}
    p^R_{ij}=\frac{1}{\sqrt{2}}(\mathbf e^{+}_{ij}+i\mathbf e^{\times}_{ij}), \ \ p^L_{ij}=\frac{1}{\sqrt{2}}(\mathbf e^{+}_{ij}-i\mathbf e^{\times}_{ij}).
\end{equation}
The plus and cross polarization tensors can be expressed as
\begin{equation}
\label{poltensor1}
\begin{split}
\mathbf e^+_{ij}=&\frac{1}{\sqrt{2}}(\mathbf e_i \mathbf e_j-\bar{\mathbf e}_i \bar{\mathbf e}_j),\\
\mathbf e_{ij}^\times=&\frac{1}{\sqrt{2}}(\mathbf e_i\bar{\mathbf e}_j+\bar{\mathbf e}_i \mathbf e_j),
\end{split}
\end{equation}
where ${\mathbf e_{i}\left(\bm{k}\right)}$ and ${\bar{\mathbf e}_{i}\left(\bm{k}\right)}$ are two basis vectors which are orthogonal to each other and perpendicular to the wave vector ${\bm{k}}$, i.e., satisfying ${\bm k}\cdot {\mathbf e}={\bm k}\cdot \bar{\mathbf e}={\mathbf e}\cdot \bar{\mathbf e}=0$ and $|{\mathbf e}|=|\bar{\mathbf e}|=1$.

The definition of the projection tensor is
\begin{equation}
\mathcal{T}^{lm}_{\ \ \ ij}s_{lm}(\bm{x},\eta)=\sum\limits_{A=R,L}\int \frac{\mathrm{d}^3\bm k}{(2\pi)^{3/2}}e^{i {\bm k} \cdot {\bm x}}p_{ij}^A p^{Alm}\tilde{s}_{lm}(\bm k,\eta), 
\end{equation}
where $\tilde{s}_{ij}$ is the Fourier transformation of the source $s_{ij}$.

The EoM's for the circularly polarized modes of the SIGWs in the Fourier space are
\begin{equation}\label{EQSIGW}
u^{A''}_{\bm{k}}(\eta)+\left(k^2-\frac{B^{A''}}{B^A}\right)u^{A}_{\bm{k}}(\eta)=4\frac{aS^{A}_{\bm{k}}(\eta)}{\sqrt{z^A}},
\end{equation}
where $u^A_{\bm{k}}=B^A h^A_{\bm{k}}$,
\begin{equation}\label{zA}
z^{A}(k,\eta)=1-\frac{k\lambda^A \vartheta(\varphi)'}{a^2(\eta)},
\end{equation}
with
\begin{equation}
\lambda^A=\left\{
\begin{array}{rcl}
1&& {A=R,}\\
-1&&{A=L.}
\end{array}\right.
\end{equation}
It is also required $z^{A}(k,\eta)>0$ to avoid the ghost field \cite{Dyda:2012rj}. In (\ref{EQSIGW}) we define
\begin{equation}
B^{A}(k,\eta)=a(\eta)\sqrt{z^A(k,\eta)},
\end{equation}
and the source is
\begin{equation}
S_{\bm k}^A=p^{A ij}\tilde{s}_{ij}(\bm k).
\end{equation}

Eq. \eqref{EQSIGW} can be solved by the method of the Green's function,
\begin{equation}\label{Eh}
h^{A}_{\bm k}\left(\eta\right)=\frac{4}{B^A(k,\eta)}\int^{\eta}\mathrm{d}\bar{\eta}~G^A_{k}\left(\eta,\bar{\eta}\right)
\frac{a\left(\bar{\eta}\right)S^A_{\bm k}\left(\bar{\eta}\right)}{\sqrt{z^A(k,\bar{\eta})}},
\end{equation}
where the Green's function $G^A_{k}\left(\eta,\bar{\eta}\right)$ satisfies the equation
\begin{equation}\label{GREEN}
G^{A''}_{\bm{k}}(\eta,\bar{\eta})+\left(k^2-\frac{B^{A''}}{B^A}\right)G^{A}_{\bm{k}}(\eta,\bar{\eta})=\delta(\eta-\bar{\eta}).
\end{equation}

For a general coupling function $\vartheta(\varphi)$, it is difficult to find the analytic solution to Eq. \eqref{GREEN}. 
For our purpose, it is possible to get an analytic solution if the coupling function takes the exponential form
\begin{equation}
\vartheta(\varphi)=\vartheta_0\exp^{\kappa\alpha\varphi},
\end{equation}
where $\alpha$ is constant. With Eq. \eqref{beq}, we can obtain 
\begin{equation}
\varphi=\frac{2\epsilon^S}{\kappa}\ln(\eta/\eta_0)+\varphi_0,
\end{equation}
where $\varphi_0$ is the value of $\varphi$ at $\eta_0$ and $\epsilon^S=\pm1$, which corresponds to $\varphi'=\pm 2/(\kappa\eta)$, respectively.

Combining the above two equations, $z^A$ can be expressed as
\begin{equation}
z^A(k,\eta)=1-\frac{k\lambda^A 2\alpha \epsilon^S\vartheta_0\text{exp}^{\kappa \alpha \varphi_0}}{a^2_0\eta^{2\alpha \epsilon^S}_0}\eta^{2\alpha \epsilon^S-3}.
\end{equation}
Generally, $z^A(k,\eta)$ depends on both $\eta$ and $k$, which makes the analysis involved. Nevertheless,
if $2\alpha \epsilon^S-3=0$, $z^A$ is independent of $\eta$, 
\begin{equation}\label{coz}
z^A(k)=1-z_0k\lambda^A,
\end{equation}
with $z_0=3\vartheta_0\text{exp}^{\kappa \alpha \varphi_0}/(a^2_0\eta^3_0)$, which enables us to solve Eq. \eqref{GREEN} analytically. In the following, we choose the value of $\alpha$ such that $2\alpha \epsilon^S-3=0$. Moreover, there is a constraint on $z_0$ to keep $z^A$ positive, that is
\begin{equation}
z_0 k< 1.
\end{equation}
We will see that this constraint makes the contribution to $\Omega_{\mathrm{GW}}$ from the CS term is at most of the same order as that from GR, but never dominates.
With the above settings, the Green's function is exactly the same as that in the standard GR,
\begin{equation}
G^{A}_{\bm{k}}(\eta,\bar{\eta})=\frac{\sin[k(\eta-\bar{\eta})]}{k}\Theta(\eta-\bar{\eta}),
\end{equation}
where $\Theta$ is the Heaviside theta function, and 
\begin{equation}\label{E7}
h^{A}_{\bm k}\left(\eta\right)=\frac{4}{a(\eta)z^A(k)}\int^{\eta}\mathrm{d}\bar{\eta}~G^A_{k}\left(\eta,\bar{\eta}\right)
a\left(\bar{\eta}\right)S^A_{\bm k}(\bar{\eta}),
\end{equation}

The source term $S^A_{\bm k}$ can be expressed as
\begin{equation}
S^A_{\bm k}=S^{A(\mathrm{scalar})}_{\bm k}+S^{A(\mathrm{PV})}_{\bm k},
\end{equation}
where 
\begin{equation} \label{srcsca}
 S^{A(\mathrm{scalar})}_{\bm k}=\int \frac{\mathrm{d}^3\bm k'}{(2\pi)^{3/2}}p^{Aij}k^{'}_i k^{'}_j \zeta(\bm k')\zeta(\bm k-\bm k ') f_{\mathrm{scalar}}(u,v,x),
\end{equation}
with
\begin{equation}
    p^{Aij}k^{'}_i k^{'}_j=k^{'2}\sin^2(\theta)\text{e}^{2i\lambda^A\ell},
\end{equation}
where $\theta$ is the angle between $\bm{k}'$ and $\bm{k}$, $\ell$ is the azimuthal angle of $\bm{k}'$, and
\begin{gather}\label{fNewton}
f_{\mathrm{scalar}}(u,v,x)=\frac{4}{9}\left(2T_\psi(vx)T_\psi(ux)+[T_\psi(vx)+vxT_\psi^{\ast}(vx)][T_\psi(ux)+uxT_{\psi}^{\ast} (ux)]\right),
\end{gather}
where $u=k'/k$, $v=|\bm k-\bm{k}'|/k$, $x=k\eta$, and the ``$\ast$'' denotes the derivative with respect to the argument. In deriving Eq.  \eqref{fNewton}, we have used the relation Eq. \eqref{1st_eq_3}, $\phi=\psi$. For later convenience, we have also symmetrized $f_{\mathrm{scalar}}(u, v, x)$ with respect to $u \leftrightarrow v$. Here, the scalar perturbations have been split into contributions from the primordial curvature perturbation and the transfer function,
\begin{gather}
\label{defs}
\delta \varphi(\bm k,\eta)=\frac{2}{3}\zeta(\bm k) kT_{\delta \varphi}(x),\\
\psi(\bm k,\eta)=\frac{2}{3}\zeta(\bm k)T_\psi(x),
\end{gather}
where $\zeta(\bm k)$ is the primordial curvature perturbation.
Since the evolution of linear scalar perturbations is unaffected by PV, the transfer function for $\psi$ during the radiation-dominated era is the same as that is in GR,
\begin{equation}\label{TN}
 T_{\psi}(x)=\frac{9}{x^2}\left(\frac{\sin(x/\sqrt{3})}{x/\sqrt{3}}-\cos(x/\sqrt{3})\right).
\end{equation}

Similarly, for the PV contribution to the source,
\begin{equation} \label{srcpv}
    S^{A(\mathrm{PV})}_{\bm k}=\int \frac{\mathrm{d}^3\bm k'}{(2\pi)^{3/2}}p^{Aij}k^{'}_i k^{'}_j \zeta(\bm k')\zeta(\bm k-\bm k ') f^A_{\mathrm{PV}}(k,u,v,x),
\end{equation}
where
\begin{equation}
   f^A_{\mathrm{PV}}(k,u,v,x)=-\frac{\lambda^A k^3}{a^2}\frac{4}{9}\partial_{x}\left[\vartheta_{\varphi}\left(uT_{\delta \varphi}(ux)T_{\psi}(vx)+vT_{\delta \varphi}(vx)T_{\psi}(ux)\right)\right],
\end{equation}
and we have symmetrized $f_{\mathrm{PV}}(u, v, x)$ under $u \leftrightarrow v$. With the expression of $\vartheta(\varphi)$, the above equation becomes
\begin{equation}
\begin{split}
f^A_{\mathrm{PV}}(k,u,v,x)=&-z_0\frac{4}{9}\frac{\kappa k^2\lambda^A \epsilon^S}{2}\left[3\left(uT_{\delta \varphi}(ux)T_{\psi}(vx)+vT_{\delta \varphi}(vx)T_{\psi}(ux)\right)\right.\\& \left.+ux\left(uT^{\ast}_{\delta \varphi}(ux)T_{\psi}(vx)+vT_{\delta \varphi}(ux)T^{\ast}_{\psi}(vx)\right) \right. \\ & \left. +vx\left(vT^{\ast}_{\delta \varphi}(vx)T_{\psi}(ux)+uT_{\delta \varphi}(vx)T^{\ast}_{\psi}(ux)\right)\right].
\end{split}
\end{equation}
With Eqs. \eqref{pvarphi}, \eqref{beq} and \eqref{TN}, we obtain the transfer function for $\delta \varphi$ to be
\begin{equation}
\begin{split}
T_{\delta \varphi}(x)&=\kappa^{-1}k^{-1}\epsilon^S\left(x T^{\ast}_{\psi}(x)+T_{\psi}(x)\right)\\&=\kappa^{-1}k^{-1}\epsilon^S\frac{3}{x^3}\left(6x\cos(x/\sqrt{3})+\sqrt{3}(-6+x^2)\sin(x/\sqrt{3})\right).
\end{split}
\end{equation}

By plugging Eqs. \eqref{srcsca} and  \eqref{srcpv} in Eq.  \eqref{E7}, the solutions of the circularly polarized modes can be written in a compact form
\begin{equation}
\label{hsolution}
h^A_{\bm k}(\eta)=\frac{4}{z^A(k)} \int\frac{\mathrm{d}^3\bm k'}{(2\pi)^{3/2}} p^{Aij}k^{'}_i k^{'}_j\zeta(\bm k')\zeta(\bm k-\bm k')\frac{1}{k^2}I^{A}(k,u,v,x),
\end{equation}
where
\begin{equation}
\begin{split}
\label{I_int}
I^{A}(k,u,v,x)&=\int_0^x\mathrm{d}\bar{x}\frac{a(\bar{\eta})}{a(\eta)}k G^A_{k}(\eta,\bar{\eta})\left(f_{\mathrm{scalar}}(u,v,\bar x)+f^A_{\mathrm{PV}}(k,u,v,\bar x)\right)\\&
= I_{\mathrm{GR}}(u,v,x)+I^{A}_{\mathrm{PV}}(k,u,v,x).
\end{split}
\end{equation}
In the above
\begin{equation}
\label{ISC}
I_{\mathrm{GR}}(u,v,x)=\int_0^x\mathrm{d}\bar{x}\frac{a(\bar{\eta})}{a(\eta)}kG_{k}(\eta,\bar{\eta})f_{\mathrm{scalar}}(u,v,\bar x),\\
\end{equation}
is the same as that in GR \cite{Kohri:2018awv}, and
\begin{equation}
\label{IPV}
I^{A}_{\mathrm{PV}}(k,u,v,x)=\int_0^x\mathrm{d}\bar{x}\frac{a(\bar{\eta})}{a(\eta)}k G_{k}(\eta,\bar{\eta})f^A_{\mathrm{PV}}(k,u,v,\bar x),
\end{equation}
is the contribution from the PV terms, of which the concrete expression can be found in appendix \ref{kernelI}.

The power spectrum of the SIGWs $\mathcal{P}_{h}^{A}$ is defined by 
\begin{equation}
\langle h^A_{\bm{k}} h^C_{\bm{k}'}\rangle =\frac{2\pi^2}{k^3}\delta^3(\bm k+\bm k')\delta^{AC}\mathcal{P}^{A}_{h}(k).
\end{equation}
With the solution to Eq. \eqref{E7} and the definition of $\mathcal{P}_{h}^{A}$, we can obtain the power spectra of the SIGWs from the CS modified gravity
\begin{align}
\label{PStensor}
\mathcal{P}^{A}_h(k,x)=4\int_{0}^\infty\mathrm{d}u\int_{|1-u|}^{1+u}\mathrm{d}v
\mathcal{J}(u,v)\frac{I^{A}(k,u,v,x)^2}{(z^A(k))^2}\mathcal{P}_\zeta(uk)\mathcal{P}_\zeta(vk),
\end{align}
where
\begin{equation}
\mathcal{J}(u,v)=\left[\frac{4u^2-(1+u^2-v^2)^2}{4uv}\right]^2,
\end{equation}
and $\mathcal{P}_\zeta$ is the power spectrum of primordial curvature perturbation.

The fractional energy density of the SIGWs is
\begin{equation}\label{OGW}
\begin{split}
\Omega_{\mathrm{GW}}(k,x)&=\frac{1}{48}\left(\frac{k}{\mathcal{H}}\right)^2\sum\limits_{A=R,L}\overline{\mathcal{P}^A_h(k,x)}=\frac{x^2}{48}\sum\limits_{A=R,L}\overline{\mathcal{P}^A_h(k,x)}\\
&=\frac{1}{12}\int_{0}^\infty\mathrm{d}u\int_{|1-u|}^{1+u}\mathrm{d}v
\mathcal{J}(u,v)\sum\limits_{A=R,L}\frac{\overline{\tilde{I}^{A}(k,u,v,x)^2}}{(z^A(k))^2}\mathcal{P}_\zeta(uk)\mathcal{P}_\zeta(vk),
\end{split}
\end{equation}
where the overline represents the time average, and $\overline{\tilde{I}^{A}(k,u,v,x)^2}=\overline{I^{A}(k,u,v,x)^2}x^2$.

The GWs behave as free radiation, thus the fractional energy density of the SIGWs at the present time $\Omega_{\mathrm{GW},0}$ can be expressed as 
\cite{Espinosa:2018eve}
\begin{equation}\label{EGW}
\Omega_{\mathrm{GW},0}\left(k\right)=\Omega_{\mathrm{GW}}\left(k,\eta\rightarrow\infty\right)\Omega_{r,0},
\end{equation}
where $\Omega_{r,0}$ is the current fractional energy density of the radiation.

For later convenience, in the following we show the explicit expression for the averaged kernel $\overline {(I^A)^2}$, of which the detailed derivation can be found in appendix \ref{kernelI}. Since we are interested in the SIGWs at the present time, we can take $x\gg 1$. In this limit, we have
\begin{equation}\label{IK}
\begin{split}
\overline{I^A(k,u,v,x\rightarrow \infty)^2}&=\frac{1}{2x^2}\left(\left(I_{\mathrm{GR}s}+I^A_{\mathrm{PV}s}\right)^2+\left(I_{\mathrm{GR}c}+I^A_{\mathrm{PV}c}\right)^2\right)\\&=\frac{\left(6(u^2+v^2-3)+3z_0\lambda^A k(u+v)(3+(u-v)^2)\right)^2}{2x^2(8u^3v^3)^2}\\& \Big[\left(-4uv+(u^2+v^2-3)\ln\Big|\frac{3-(u+v)^2}{3-(u-v)^2}\Big|\right)^2 \\& +\left(u^2+v^2-3\right)^2\pi^2 \Theta\left(u+v-\sqrt{3}\right) \Big],
\end{split}
\end{equation}
with $z_0k<1$. From the above expression, the magnitude of the contribution from the PV term can be at most of the same order as that from the GR.

The degree of the circular polarization is defined as \cite{Saito:2007kt,Gluscevic:2010vv}
\begin{equation}\label{pi}
\Pi\equiv\frac{\overline{\mathcal{P}^R_h}-\overline{\mathcal{P}^L_h}}{\overline{\mathcal{P}^R_h}+\overline{\mathcal{P}^L_h}},
\end{equation}
where again an overline denotes the time average.
Combining Eqs. \eqref{pi} and \eqref{avi}, we obtain
\begin{equation}
\Pi=\frac{\int_{0}^\infty\mathrm{d}u\int_{|1-u|}^{1+u}\mathrm{d}v
\mathcal{J}(u,v)\mathcal{N}(k,u,v)\mathcal{P}_\zeta(uk)\mathcal{P}_\zeta(vk)}{\int_{0}^\infty\mathrm{d}u\int_{|1-u|}^{1+u}\mathrm{d}v\mathcal{J}(u,v)\mathcal{M}(k,u,v)\mathcal{P}_\zeta(uk)\mathcal{P}_\zeta(vk)},
\end{equation}
where
\begin{equation}\label{nN}
\mathcal{N}(k,u,v)=\frac{\overline{I^R(k,u,v,x\rightarrow \infty)^2}}{(z^R(k))^2}-\frac{\overline{I^L(k,u,v,x\rightarrow \infty)^2}}{(z^L(k))^2},
\end{equation}
and
\begin{equation}\label{mM}
\mathcal{N}(k,u,v)=\frac{\overline{I^R(k,u,v,x\rightarrow \infty)^2}}{(z^R(k))^2}+\frac{\overline{I^L(k,u,v,x\rightarrow \infty)^2}}{(z^L(k))^2},
\end{equation}
with $I^A=I_{\mathrm{GR}}+I^A_{\mathrm{PV}}$. According to Eq. \eqref{IK} and the fact that $I^R_{\mathrm{PV}}=-I^L_{\mathrm{PV}}$, the degree of the circular polarization can be large only if $\mathcal{O}\left(I_{\mathrm{GR}}\right)\sim \mathcal{O}(I^A_{\mathrm{PV}})$, which is possible as long as $\mathcal{O}\left(2(u^2+v^2-3)\right)\sim \mathcal{O}(z_0 k(u+v)(3+(u-v)^2))$.

\section{Examples with concrete forms for the power spectrum of the curvature perturbation}\label{IV}

The expressions derived in the above section are quite general, which in fact can be applied to more general gravity theories that are different from GR.
In order to investigate the features of SIGWs from the CS modified gravity, in this section we consider some phenomenological forms for the power spectrum of the curvature perturbation that are commonly used to study the SIGWs. 

\subsection{The monochromatic power spectrum}

First we consider the curvature perturbation with the $\delta$-function-type power spectrum \cite{Ananda:2006af,Inomata:2016rbd}
\begin{equation}\label{ps1}
    \mathcal{P}_\zeta(k)=\mathcal{A}_\zeta\delta(\ln(k/k_p)).
\end{equation}
After some straightforward calculations, we obtain the fractional energy density of the SIGWs
\begin{equation}\label{ph}
\Omega_{\mathrm{GW}}(k)=\frac{\mathcal{A}^2_\zeta \bar{k}^{-2}}{12}\left(\frac{4-\bar{k}^2}{4}\right)^2\sum\limits_{A=R,L}\frac{\overline{\tilde{I}^A(k,\bar{k}^{-1},\bar{k}^{-1},x\rightarrow \infty)^2}}{(1-z_0\lambda^Ak)^2}\Theta(2-\bar{k}),
\end{equation}
where $\bar{k}=k/k_p$.

\begin{figure}[htp]
\centering
\subfigure{\includegraphics[width=0.45\linewidth]{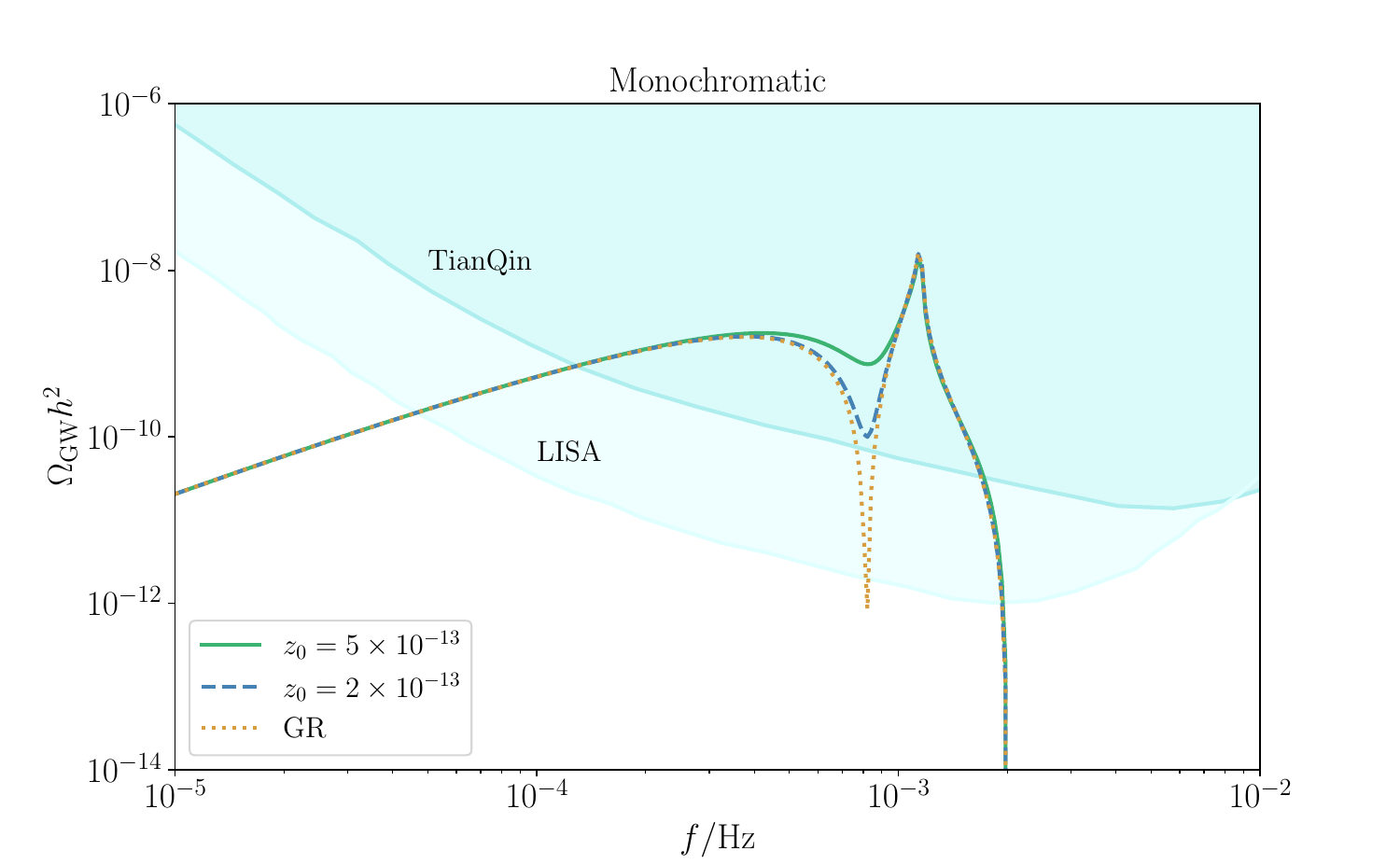}}
\subfigure{\includegraphics[width=0.45\linewidth]{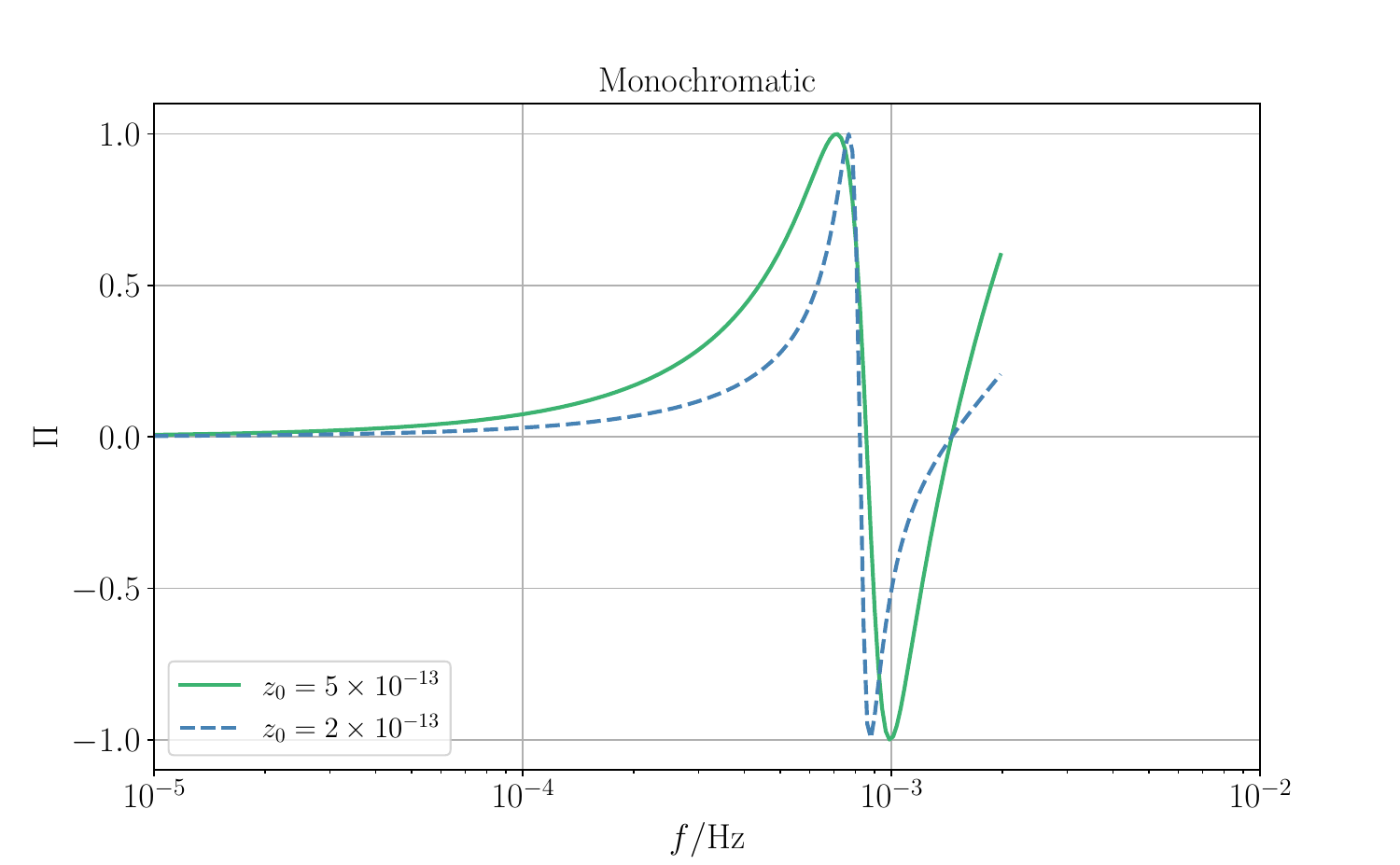}}
\caption{The energy density of SIGWs from GR and CS gravity (left panel) and the degree of the circular polarization of SIGWs from CS term (right panel). The peak amplitude and the peak scale are fixed to be $\mathcal{A}_{\zeta}=10^{-2}$ and $k_p=10^{12}\text{Mpc}^{-1}$, respectively, which corresponds to the maximum sensitivity of TianQin and LISA.}
\label{fig:GWs1}
\end{figure}

With the expression of $\overline{(I^A)^2}$, i.e., Eqs. \eqref{EGW} and \eqref{ph}, we plot the fractional energy density of the SIGWs from GR and CS modified gravity in the left panel of Fig. \ref{fig:GWs1}, respectively. 

From Fig. \ref{fig:GWs1}, there is a divergence at the frequency $\bar{f}=f/f_p=2/\sqrt{3}$ due to the resonant amplification \cite{Ananda:2006af,Kohri:2018awv}. We can also see that from the expression of $I^A$, the divergence comes from the cosine integral $\text{Ci}$. There is a term in $I^A_{s}$ that is $\propto (2\bar{f}^{-2}-3)^2\text{Ci}(|1-2\bar{f}^{-1}/\sqrt{3}|)$, which is divergent when $\bar{f}=2/\sqrt{3}$. From Fig. \ref{fig:GWs1}, we can see that as $k$ increases, the contribution from the PV term to $\Omega_{\mathrm{GW}}$ becomes comparable with the contribution from GR, but the former never dominates due to the constraint \eqref{coz}, i.e., $z_0k<1$.

After some manipulations, we obtain
\begin{equation}
\Pi=\frac{\mathcal{N}(k,\bar{k}^{-1},\bar{k}^{-1})}{\mathcal{M}(k,\bar{k}^{-1},\bar{k}^{-1})}\Theta(2-\bar{k}).
\end{equation}
The degree of the circular polarization of the SIGWs is tiny, i.e., $|\Pi| \ll 1$, when the deviation from GR is negligible, i.e.,  when the contribution from the CS term  $I^A_{\mathrm{PV}}$ is subdominant. Only in the case that the contribution to $\Omega_{\mathrm{GW}}$ attributed to the CS term is about the same order as that from GR, namely $\mathcal{O}(I_{\mathrm{GR}}) \sim \mathcal{O}(I^A_{\mathrm{PV}})$, we may obtain a large $\Pi$. 

At lower frequencies, the contribution to the degree of the circular polarization $\Pi$ from the CS term is negligible, i.e., $|\Pi|\ll 1$, which is mainly because the extra term $z_0k\ll1$ in $I^A_{\mathrm{PV}}$. As the frequency increases, the contribution to $\Omega_{\mathrm{GW}}$ attributed to the CS term becomes comparable with that from GR gradually, and thus $|\Pi|$ becomes large. When $z_0 k \sim \mathcal{O}(1)$, i.e, at frequency $f\simeq 10^{-3}$Hz, we obtain the maximum of the degree of the circular polarization, $|\Pi|\simeq 1$, as shown in Fig. \ref{fig:GWs1}.

\subsection{The lognormal peak}

\begin{figure}[htp]
\centering
\subfigure{\includegraphics[width=0.45\linewidth]{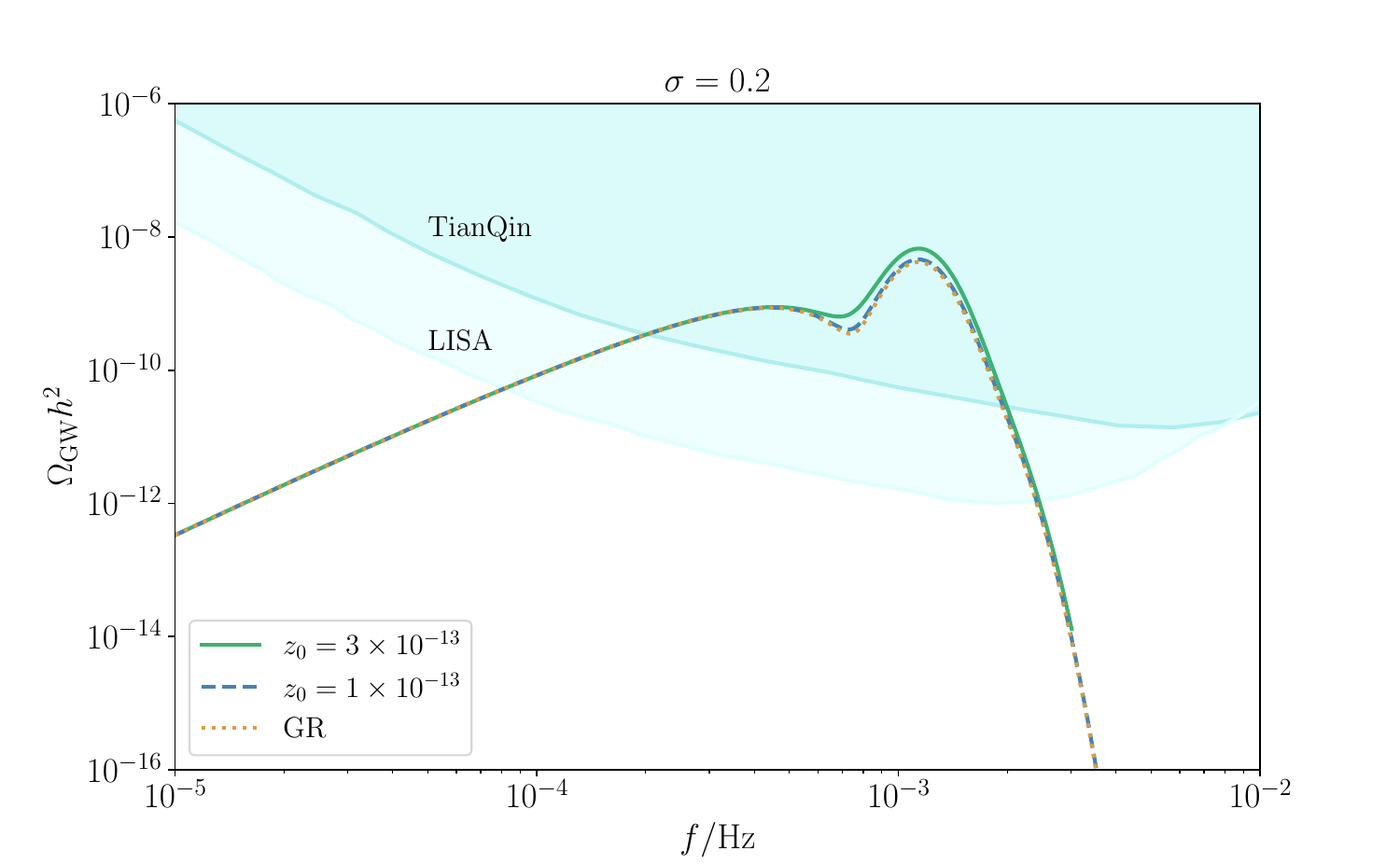}}
\subfigure{\includegraphics[width=0.45\linewidth]{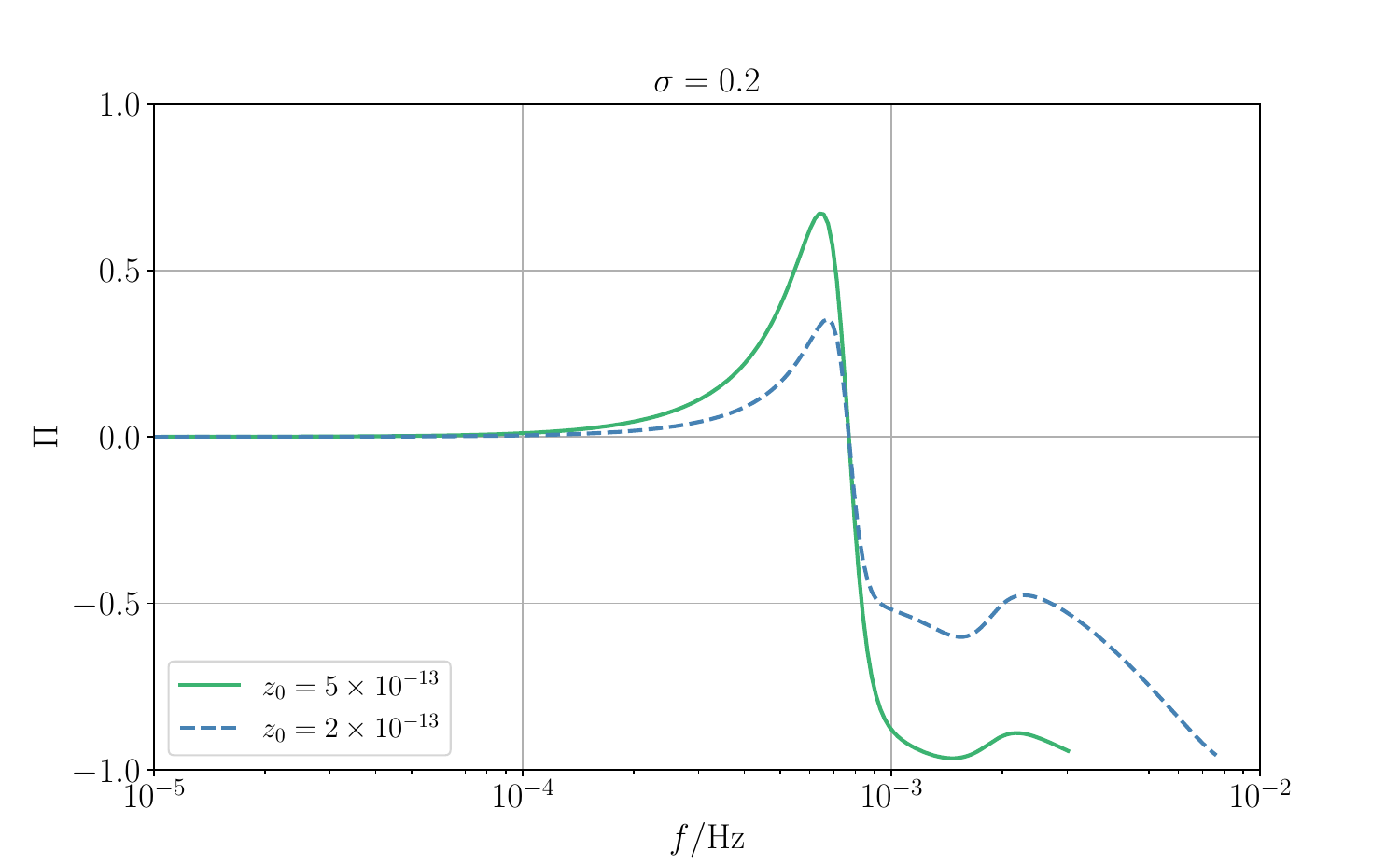}}
\caption{The SIGWs induced by the lognormal peak with $\sigma = 0.2$. We plot the contributions to the energy density of SIGWs from the GR and the CS term (left panel) and the degree of the circular polarization of SIGWs from the CS gravity (right panel). The parameters $\mathcal{A}_{\zeta}$ and $k_p$ are the same as those in the case of monochromatic power spectrum.}
\label{fig:GWs2}
\end{figure}

\begin{figure}[htp]
\centering
\subfigure{\includegraphics[width=0.45\linewidth]{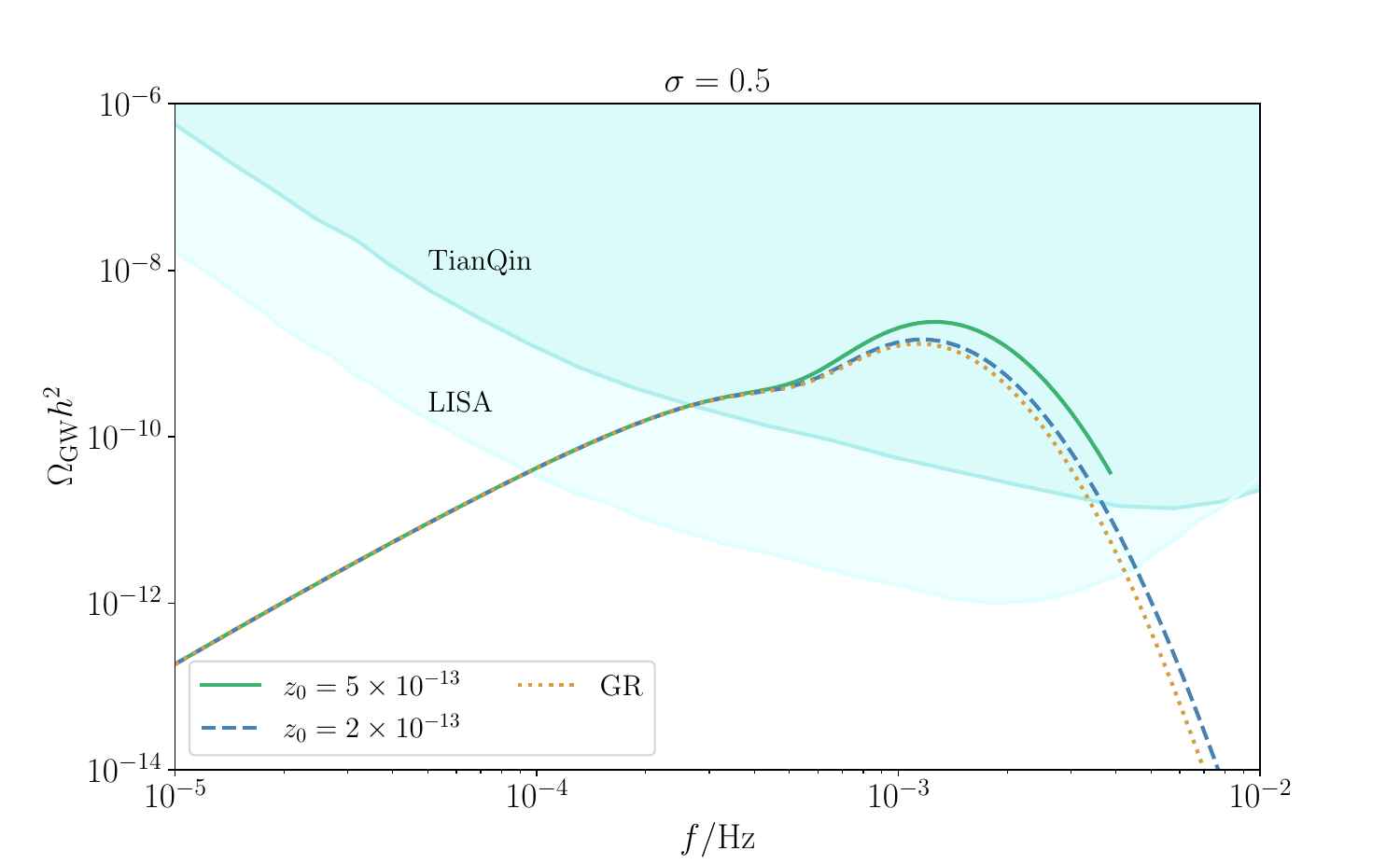}}
\subfigure{\includegraphics[width=0.45\linewidth]{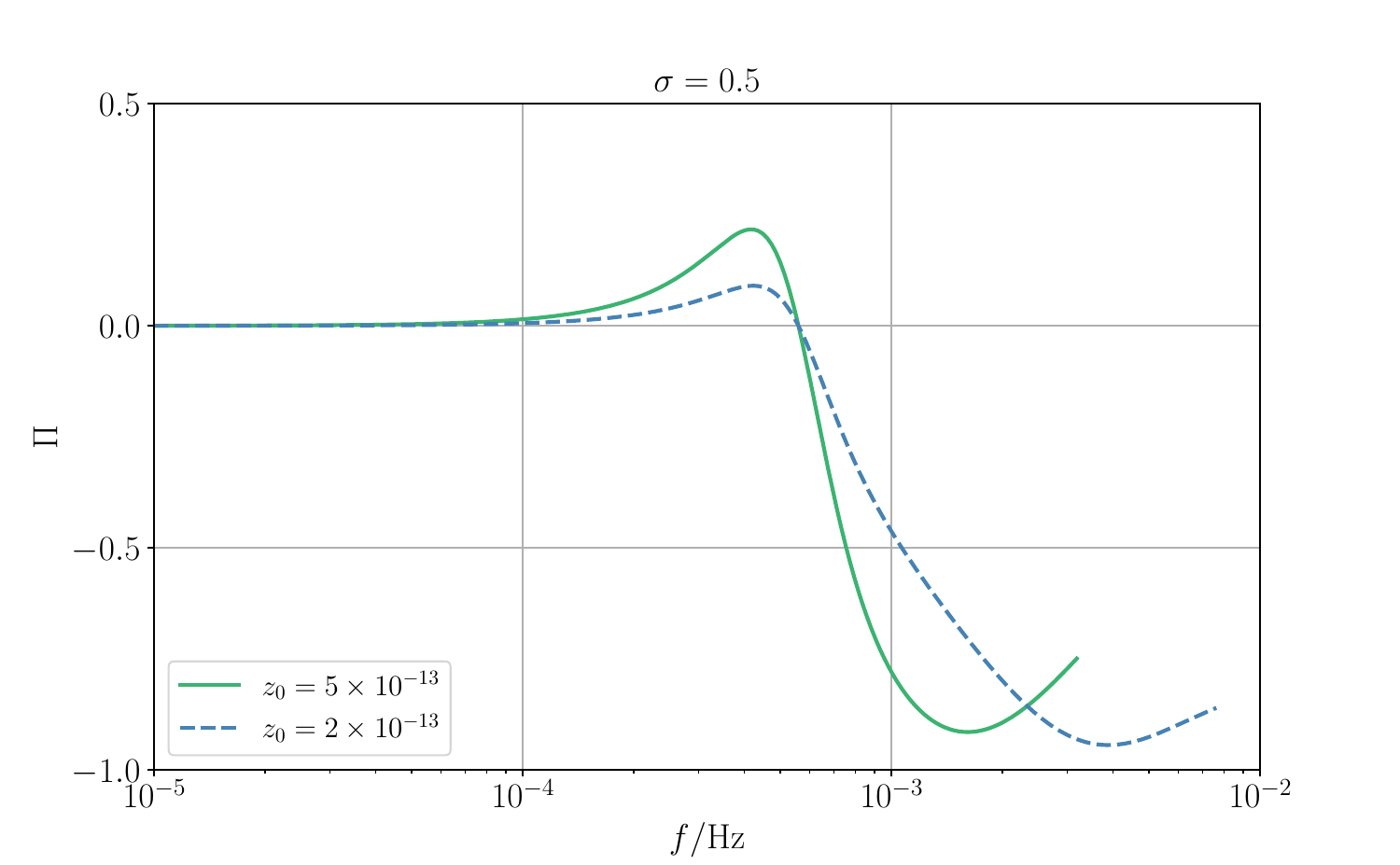}}
\caption{The SIGWs induced by the lognormal peak with $\sigma = 0.5$. We plot the contributions to the energy density of SIGWs from the GR and the CS term  (left panel) and the degree of the circular polarization of SIGWs from the CS gravity (right panel).  The parameters $\mathcal{A}_{\zeta}$ and $k_p$ are the same as those in the case of monochromatic power spectrum.}
\label{fig:GWs3}
\end{figure}

We consider another power spectrum for the curvature perturbation with Gaussian form \cite{Inomata:2018epa,Pi:2020otn}
\begin{equation}\label{ps2}
\mathcal{P}_{\zeta}(k)=\frac{\mathcal{A}_{\zeta}}{\sqrt{2\pi}\sigma}\text{exp}^{-\frac{\ln^2(k/k_p)}{2\sigma^2}}.
\end{equation}
The above power spectrum becomes monochromatic Eq.  \eqref{ps1} when $\sigma\rightarrow 0$. Unlike the monochromatic power spectrum, the SIGWs generated by this power spectrum shows no divergence as long as $\sigma$ is not very small. We have numerically evaluated the fractional energy density and the degree of the circular polarization of the SIGWs from the CS modified gravity generated by the above lognormal peak. 
The results are shown in Fig. \ref{fig:GWs2} and Fig. \ref{fig:GWs3} for $\sigma=0.2$ and $\sigma=0.5$, respectively. 

It is clear that from Fig. \ref{fig:GWs2} and Fig. \ref{fig:GWs3}, the resonant divergence disappears since more modes with a wider range of wavenumbers contribute to the integration Eq. \eqref{PStensor}. Similar to the previous subsection, the degree of the circular polarization can be large only if the magnitude of the energy density attributed to the CS term is about the same order as that from GR, namely $\mathcal{O}(I_{\mathrm{GR}}) \sim \mathcal{O}(I^A_{\mathrm{PV}})$. 
From Figs. \ref{fig:GWs2} and \ref{fig:GWs3}, we can obtain a large degree of circular polarization $|\Pi|\simeq 1$, although the correction to $\Omega_{\mathrm{GW}}$ from the parity violating term is at most of the same order as that from the GR. The reason is the same as what we have explained in the above subsection. 

\section{Conclusion}\label{V}

In this paper, we made the first step to use SIGWs to test the possible parity violation in the early universe and/or in the gravitational interactions. We took the simplest CS modified gravity with a dynamical scalar field Eq. \eqref{action} as the example.  
We derived the EoMs for the SIGWs, and by taking the coupling to  be $\vartheta(\varphi)=\vartheta_0\exp^{\kappa\alpha \varphi}$, we gave the semianalytic expressions that can be used to evaluate the effects on the SIGWs from the CS gravity.
The general equations for the circular polarization modes of the SIGWs are given in Eq. \eqref{EQSIGW}. Generally, the SIGWs receive contributions from two parts, one is the same as in the case of GR Eq. \eqref{srcsca}, the other is the correction due to the presence of CS term Eq. \eqref{srcpv}. Interestingly, we found that the contribution to $\Omega_{\mathrm{GW}}$ from the PV term is at most of the same order as that from GR, and thus will never dominate. On the contrary, the degree of circular polarization of SIGWs can be as large as $|\Pi|\simeq 1$. This fact implies that even the contribution to the energy density of the SIGWs from the PV term is a small correction, the signal of the SIGWs can be significantly polarized.

In order to illustrate the features of the SIGWs due to the CS term, in Sec. \ref{IV}
we evaluate the fractional energy density $\Omega_{\mathrm{GW}}$ with the monochromatic and lognormal power spectra of primordial curvature perturbation. We also evaluate the degree of the circular polarization of SIGWs, $\Pi$ defined in Eq. \eqref{pi}. We found that $\Pi$ becomes large only when the contributions to $\Omega_{\mathrm{GW}}$ from the GR and the CS term are comparable to each other. The results present in this paper indicate that SIGWs may provide us a new tool to test cosmological models with parity violation.

\begin{acknowledgments}
This work was partly supported by the National Natural Science Foundation of China (NSFC) under the grant No. 11975020. 
The equations of motion for the SIGWs from CS gravity are derived with the help of the Mathematica package \textit{xPand} \cite{Pitrou:2013hga}.
\end{acknowledgments}

\appendix

\section{The integral kernel}\label{kernelI}

For the sake of clarity, we split $I^A$ defined in Eq. \eqref{I_int} into two parts as follows
\begin{equation}
I^A(k,u,v,x)=\frac{\sin(x)}{x}\left(I_{\mathrm{GR}s}+I^A_{\mathrm{PV}s}\right)+\frac{\cos(x)}{x}\left(I_{\mathrm{GR}c}+I^A_{\mathrm{PV}c}\right),
\end{equation}
where the subscript ``s'' and ``c'' stand for contributions involving the sine and cosine functions, respectively. We also write
\begin{gather}
 I^{A}_{\mathrm{PV}s}(k,u,v,x)=\mathcal{I}^A_{\mathrm{PV}s}(k,u,v,x)- \mathcal{I}^A_{\mathrm{PV}s}(k,u,v,0), \nonumber\\
 I^{A}_{\mathrm{PV}c}(k,u,v,x)=\mathcal{I}^A_{\mathrm{PV}c}(k,u,v,x)- \mathcal{I}^A_{\mathrm{PV}c}(k,u,v,0),
\end{gather}
where $\mathcal{I}^{A}_{\mathrm{PV}s}$ and $\mathcal{I}^{A}_{\mathrm{PV}c}$ are defined by
\begin{gather}
\mathcal{I}^{A}_{\mathrm{PV}s}(k,u,v,y)=\int \mathrm{d} y \cos y f_{\mathrm{PV}}(k,u,v,y) y, \nonumber\\
\mathcal{I}^{A}_{\mathrm{PV}c}(k,u,v,y)=-\int \mathrm{d} y \sin y f_{\mathrm{PV}}(k,u,v,y) y.
\end{gather}

After tedious manipulations, the concrete expressions of $\mathcal{I}^A_{\mathrm{PV}s}$ and $\mathcal{I}^A_{\mathrm{PV}c}$ are found to be
\begin{align}
\mathcal{I}^{A}_{\mathrm{PV}s}(k,u,v,y)=&\frac{3z_0\lambda^A k(u+v)}{8 u^3 v^3 y^4}  \Big\{72uvy^2\cos y \cos\frac{uy}{\sqrt{3}}\cos\frac{v y}{\sqrt{3}}-24uvy^3\cos\frac{uy}{\sqrt{3}}\cos\frac{vy}{\sqrt{3}}\sin y \nonumber\\& +4\sqrt{3}vy\left(-18+(3+3u^2-2uv+v^2)y^2\right)\cos y \cos \frac{vy}{\sqrt{3}}\sin\frac{uy}{\sqrt{3}}\nonumber\\&+4\sqrt{3}uy\left(-18+(3+3v^2-2uv+u^2)y^2\right)\cos y \cos \frac{uy}{\sqrt{3}}\sin\frac{vy}{\sqrt{3}} \nonumber\\& +24\sqrt{3}vy^2\cos \frac{vy}{\sqrt{3}} \sin y \sin \frac{uy}{\sqrt{3}} 
+24\sqrt{3}uy^2\cos \frac{uy}{\sqrt{3}} \sin y \sin \frac{vy}{\sqrt{3}}\nonumber\\& -12\left(-18+(3+3u^2-2uv+3v^2)y^2\right)\cos y\sin\frac{uy}{\sqrt{3}} \sin\frac{vy}{\sqrt{3}}\nonumber\\&+12y\left(-6+(3+u^2-2uv+v^2)y^2\right)\sin y\sin\frac{uy}{\sqrt{3}}\sin\frac{vy}{\sqrt{3}}\Big\}\nonumber\\&
-\frac{3z_0\lambda^A k(u+v)}{8 u^3 v^3}\left(-9+u^4-2u^3v+2u^2v^2+v^4+6uv-2uv^3\right)\nonumber\\&\left(\text{Ci}\left[\left(1+\frac{u+v}{\sqrt{3}}\right)y\right]+\text{Ci}\left[\left|1-\frac{u+v}{\sqrt{3}}\right|y\right]\right.\nonumber\\&\left.-\text{Ci}\left[\left(1+\frac{u-v}{\sqrt{3}}\right)y\right]-\text{Ci}\left[\left(1-\frac{u-v}{\sqrt{3}}\right)y\right]\right),
\end{align}

\begin{align}
\mathcal{I}^{A}_{\mathrm{PV}c}(k,u,v,y)=&\frac{3z_0\lambda^A k(u+v)}{8 u^3 v^3 y^4}  \Big\{-72uvy^2\cos\frac{uy}{\sqrt{3}}\cos\frac{v y}{\sqrt{3}}\sin y-24uvy^3\cos y\cos\frac{uy}{\sqrt{3}}\cos\frac{vy}{\sqrt{3}} \nonumber\\& +24\sqrt{3}vy^2 \cos y\cos \frac{vy}{\sqrt{3}} \sin \frac{uy}{\sqrt{3}} +24\sqrt{3}uy^2\cos y\cos \frac{uy}{\sqrt{3}}  \sin \frac{vy}{\sqrt{3}}\nonumber\\&-4\sqrt{3}vy\left(-18+(3+3u^2-2uv+v^2)y^2\right) \cos \frac{vy}{\sqrt{3}}\sin y\sin\frac{uy}{\sqrt{3}}\nonumber\\&-4\sqrt{3}uy\left(-18+(3+3v^2-2uv+u^2)y^2\right) \cos \frac{uy}{\sqrt{3}}\sin y\sin\frac{vy}{\sqrt{3}} \nonumber\\&  +12\left(-18+(3+3u^2-2uv+3v^2)y^2\right)\sin y\sin\frac{uy}{\sqrt{3}} \sin\frac{vy}{\sqrt{3}}\nonumber\\&+12y\left(-6+(3+u^2-2uv+v^2)y^2\right)\cos y\sin\frac{uy}{\sqrt{3}}\sin\frac{vy}{\sqrt{3}}\Big\}\nonumber\\&
+\frac{3z_0\lambda^A k(u+v)}{8 u^3 v^3}\left(-9+u^4-2u^3v+2u^2v^2+v^4+6uv-2uv^3\right)\nonumber\\&\left(\text{Si}\left[\left(1+\frac{u+v}{\sqrt{3}}\right)y\right]+\text{Si}\left[\left(1-\frac{u+v}{\sqrt{3}}\right)y\right]\right.\nonumber\\&\left.-\text{Si}\left[\left(1+\frac{u-v}{\sqrt{3}}\right)y\right]-\text{Si}\left[\left(1-\frac{u-v}{\sqrt{3}}\right)y\right]\right),
\end{align}
where
\begin{equation}
\text{Si}(x)=\int_0^x \mathrm{d} y\frac{\sin y}{y},\quad \text{Ci}(x)=-\int_x^\infty \mathrm{d} y \frac{\cos y}{y}.
\end{equation}

With the above expressions, we can obtain $I^A$
\begin{equation}
I^A_s(k,u,v,x)|_{x\rightarrow\infty}=\left(I_{\text{GR}s}(u,v,x)+I^A_{\text{PV}s}(k,u,v,x)\right)|_{x\rightarrow\infty},
\end{equation}
\begin{equation}
I^A_c(k,u,v,x)|_{x\rightarrow\infty}=\left(I_{\text{GR}c}(u,v,x)+I^A_{\text{PV}c}(k,u,v,x)\right)|_{x\rightarrow\infty}.
\end{equation}
where
\begin{gather}\label{Ipv}
I^A_{\mathrm{PV}s}(k,u,v,x)|_{x\rightarrow\infty}=z_0\lambda^A k\frac{3(u+v)\left(3+(u-v)^2\right)}{8u^3v^3}\left(-4uv+(u^2+v^2-3)\ln\Big|\frac{3-(u+v)^2}{3-(u-v)^2}\Big|\right),\\
I^A_{\mathrm{PV}c}(k,u,v,x)|_{x\rightarrow\infty}=-z_0\lambda^A k\frac{3(u+v)\left(3+(u-v)^2\right)}{8u^3v^3}\left(u^2+v^2-3\right)\pi \Theta(u+v-\sqrt{3}),
\end{gather}
and \cite{Kohri:2018awv}
\begin{gather}\label{Igr}
I_{\mathrm{GR}s}(k,u,v,x)|_{x\rightarrow\infty}=\frac{3(u^2+v^2-3)}{4u^3v^3}\left(-4uv+(u^2+v^2-3)\ln\Big|\frac{3-(u+v)^2}{3-(u-v)^2}\Big|\right),\\
I_{\mathrm{GR}c}(k,u,v,x)|_{x\rightarrow\infty}=-\frac{3(u^2+v^2-3)^2}{4u^3v^3}\pi \Theta(u+v-\sqrt{3}).
\end{gather}
Combining Eqs. \eqref{Ipv} and \eqref{Igr}, we get the following results
\begin{equation}
\begin{split}
I^A_{s}(k,u,v,x)|_{x\rightarrow\infty}=&\frac{6(u^2+v^2-3)+3z_0\lambda^A k(u+v)\left(3+(u-v)^2\right)}{8u^3v^3} \\& \times \left(-4uv+(u^2+v^2-3)\ln\Big|\frac{3-(u+v)^2}{3-(u-v)^2}\Big|\right),
\end{split}
\end{equation}
\begin{equation}
\begin{split}
I^A_{c}(k,u,v,x)|_{x\rightarrow\infty}=&-\frac{6(u^2+v^2-3)+3z_0\lambda^A k(u+v)\left(3+(u-v)^2\right)}{8u^3v^3}\\& \times (u^2+v^2-3)\pi \Theta(u+v-\sqrt{3}).
\end{split}
\end{equation}

As a result, the time average 
\begin{equation}\label{avi}
\overline{I^A(k,u,v,x\rightarrow \infty)^2}=\frac{1}{2x^2}\left(I^A_{s}(k,u,v,x\rightarrow \infty)^2+I^A_{c}(k,u,v,x\rightarrow \infty)^2\right).
\end{equation}


%

\end{document}